\documentclass[a4paper,11pt]{article}
\pdfoutput=1
\usepackage{jheppub}
\usepackage{braket}
\usepackage{bm}
\usepackage{subcaption}
\usepackage{todonotes}
\renewcommand{\Re}{\operatorname{Re}}

\title{$\mathcal{O}\left(m\alpha^2 (Z\alpha)^6\right)$ contribution to Lamb shift from radiative corrections to the Wichmann-Kroll potential.}

\author{Petr A. Krachkov}
\author{and Roman N. Lee}

\affiliation{Budker Institute of Nuclear Physics, Novosibirsk 630090, Russia}

\emailAdd{p.a.krachkov@inp.nsk.su}
\emailAdd{r.n.lee@inp.nsk.su}

\abstract{
We derive an analytical expression for the contribution of the order $m\alpha^2 (Z\alpha)^6$ to the hydrogen Lamb shift which comes from the diagrams for radiative corrections to the Wichmann-Kroll potential. We use modern methods of multiloop calculations, based on IBP reduction, DRA method and differential equations.
}

\begin{document}

\maketitle

\flushbottom

\section{Introduction}

Recent advances in spectroscopy of ordinary hydrogen \cite{Bezginov2019,Grinin2020}  and deuterium \cite{PhysRevLett.104.233001}  and in their muonic analogs \cite{antognini2013proton,pohl2016laser} as well as in the electron-proton scattering \cite{xiong2019small} provide new opportunities to perform subtle tests of the bound-state quantum electrodynamics (QED).


The hydrogen atom is a system in which the role of the relativistic, radiative, and recoil effects can be investigated with high precision, both experimentally and theoretically. The corresponding corrections to energy levels are small compared to the leading term; however, with a present accuracy achieved in measuring the frequency of specific hydrogenic transitions, those contributions are large compared with the experimental uncertainty.

One of the most important bound-state QED effects is the Lamb shift. It is responsible for the $2 s_{1/2}$ – $2 p_{1/2}$ energy splitting, which in the Dirac equation approximation would be zero. The calculations of various contributions to the Lamb shift have a long history starting from Refs. \cite{Bethe1947,Karplus1951,Kroll1951,Karplus1952}, see also review \cite{Eides2007} and references therein. For the $s$-states all corrections have been calculated up to the order $m \alpha^2(Z\alpha)^6 \ln\left(1/(Z\alpha)^2\right)$. The $m \alpha^2(Z\alpha)^6$ contribution has not yet been calculated, although this correction may be important already in the next series of spectroscopic measurements.

In the present paper we calculate one of the previously unknown corrections to the Lamb shift of order $m \alpha^2(Z\alpha)^6$, which is connected with the radiative corrections to the Wichmann-Kroll (WK) potential. We use modern multiloop methods and obtain analytic result in terms of conventional polylogarithmic constants. Recently, we applied similar approach to the calculation of certain two-loop corrections to Lamb shift and hyperfine splitting in hydrogen, Ref. \cite{Krachkov:2023tly}.
The present calculation provides yet another example of the effectiveness of multiloop methods for obtaining analytic results in atomic physics.

\section{Energy shift due to radiative correction to WK potential}

\begin{figure}[h]
	\begin{subfigure}{\textwidth}
		\includegraphics[width=0.9\linewidth]{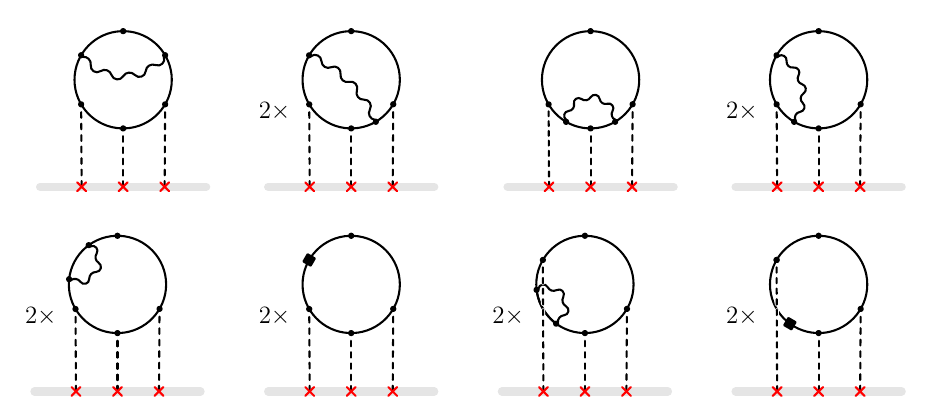}
		\caption{Diagrams with one electron loop. Black squares correspond to the mass counter term $i\delta m$.}
		\label{fig:ph}
	\end{subfigure}
	\begin{subfigure}{\textwidth}
		\includegraphics[width=0.9\linewidth]{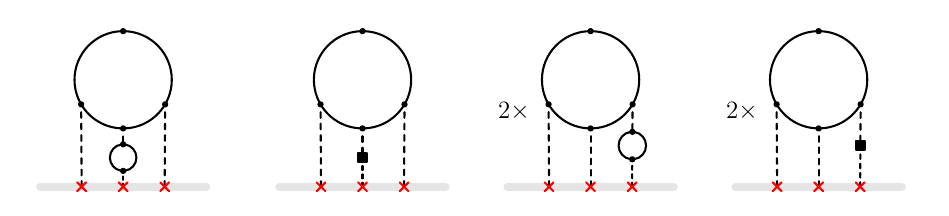}\\
		\centering\includegraphics[width=0.5\linewidth]{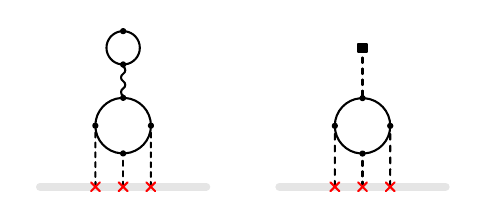}
		\caption{Diagrams with two electron loops. Black squares correspond to the vertex $-i\delta Z_A (q^2 g_{\mu\nu}-q_\mu q_\nu)$.}
		\label{fig:po}
	\end{subfigure}
	\caption{Feynman diagrams corresponding to the radiative corrections to the Wichmann-Kroll charge density $\delta\tilde{\rho}(q)$.
	}
	\label{fig:diagrams}
\end{figure}

Feynman diagrams for the radiative correction $\delta\tilde{\rho}(q)$ to the Wichmann-Kroll charge density are depicted in Fig.~\ref{fig:diagrams}. On the same figure we also show the diagrams with one-loop conterterms. The black squares on solid lines correspond to the mass counter terms $i\delta m=i m \frac{(4 \pi  \alpha ) m^{-2\epsilon}(3-2 \epsilon ) \Gamma (\epsilon )}{(4 \pi )^{2-\epsilon } (1-2 \epsilon )}$, while the black squares on dashed lines correspond to $-i\delta Z_A (q^2 g_{\mu\nu}-q_\mu q_\nu)= i \frac{(16 \pi  \alpha ) m^{-2\epsilon}\Gamma (\epsilon )}{3(4 \pi )^{2-\epsilon }}(q^2 g_{\mu\nu}-q_\mu q_\nu)$. In  principle, we should also account for the diagrams with one-loop fermion field renormalization and vertex counterterms, but their contributions cancel due to the Ward identity.

The corresponding correction to the potential $\delta \widetilde{V}(q)=\delta\tilde{\rho}(q)\cdot(e/q^2)$ contributes to energy shifts.
The characteristic atomic momenta $q$ are small compared to the electron mass. Therefore, we need the small-$q$ asymptotics of $\tilde{\rho}(q)$. This asymptotics can be analyzed using the expansion by regions approach. There are two regions which are relevant to this asymptotics. The hard region corresponds to all loop momenta $\sim m$. In this region we expand the integrand of $\delta\tilde{\rho}(q)$ in Taylor series in $q$. The zeroth term corresponds to the sum of diagrams in Fig. \ref{fig:diagrams} at $q=0$. It is easy to see that, due to the identity
\begin{multline}
	\sum_{k=1}^{N}
	\operatorname{Tr}\Big[\gamma^{\mu_1}(\hat{p}-\hat{l}_1-m)^{-1}\ldots
	\gamma^{\mu_k}(\hat{p}-\hat{l}_k-m)^{-1}\gamma^{\alpha}(\hat{p}-\hat{l}_k-m)^{-1}
	\ldots\gamma^{\mu_N}(\hat{p}-\hat{l}_N-m)^{-1}\Big]\\
	=	-\frac{\partial}{\partial p_\alpha}\operatorname{Tr}\left[\gamma^{\mu_1}(\hat{p}-\hat{l}_1-m)^{-1}\ldots  \gamma^{\mu_N}(\hat{p}-\hat{l}_N-m)^{-1}\right]\,,
\end{multline}
this sum can be written as the integral of total derivative which is zero in dimensional regularization. The sum in the left-hand side of the above identity corresponds to all possible insertions of the vertex $\gamma^{\alpha}$ in the fermion loop.

The contribution to $\delta\tilde{\rho}(q)$ linear in $\boldsymbol{q}$ is zero due to rotation symmetry. Therefore, the expansion in the hard region starts from $q^2$ term,
\begin{equation}
\delta\tilde{\rho}(q)\approx(\delta \widetilde{V}(0)/e)\cdot q^2. \label{eq:rho_expansion}
\end{equation}
Note that the sum of the two diagrams on the last row of Fig. \ref{fig:po} is suppressed by an additional factor $q^2$, therefore, they can be neglected within our present accuracy.

There is also a soft region, corresponding to all momentum transfers to the nucleus $q_1,q_2,q_3=q-q_1-q_2$ being small. Due to the gauge invariance of the light-by-light block it is easy to see that the corresponding contribution starts from $q^{4-4\epsilon}$, which is also too small for our present accuracy.

Eq. \eqref{eq:rho_expansion} shows that the potential $V(\boldsymbol{r})=e\int  e^{i\boldsymbol{q}\boldsymbol{r}}\delta\tilde{\rho}/q^2d \boldsymbol{q}/(2\pi)^3$ is proportional to the delta function and the Lamb shift contribution can be written  as:
\begin{equation}
\delta E=|\psi_{n\ell}(0)|^2\delta \widetilde{V}(0)=m\frac{\alpha^2 (Z\alpha)^6}{\pi^2 n^3}B
\delta_{\ell,0}\,,\label{eq:deltaE}
\end{equation}
where $B$ is a numerical coefficient to be calculated, $m$ is the electron mass, $n$ and $\ell$ are the principal and angular quantum numbers, respectively.
%
\section{Calculation and result}

\begin{figure}[h]
	\centering
	\includegraphics[width=0.3\linewidth]{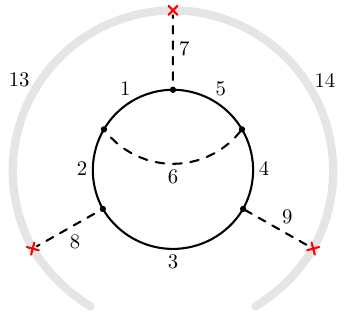}
	\caption{Topology corresponding to the integral family in Eq. \eqref{eq:j}. Numbers correspond to the subscript $k$ of the denominator $D_k$ in Eq. \eqref{eq:Ds}.}
	\label{fig:top}
\end{figure}

The small-$q$ expansion of the diagrams in Fig. \ref{fig:diagrams} can be expressed in terms of the integrals of the family
\begin{equation}
\label{eq:j}
j(n_{1},\cdots,n_{9})=\int\frac{dq_1dq_2 dl_1 dl_2}{\pi^{2d}}
\prod_{k=1}^{12}D_{k}^{-n_{k}}\times
\prod_{s=13}^{14}\frac{\delta^{(n_{s}-1)}\left(D_{s}\right)}{(n_{s}-1)!}\,,
\end{equation}
where
\begin{align}
&D_1=1-l_1^2\,,\quad D_2=1-l_2^2\,,\quad D_3=1-(l_2-q_2)^2\,,\nonumber\\
&D_4=1-(l_2+q_1)^2\,,\quad D_5=1-(l_1+q_1)^2\,,\quad D_6=-(l_1-l_2)^2\,,\nonumber\\
& D_7=-q_1^2\,,\quad D_8=-q_2^2\,,\quad D_9=-(q_1+q_2)^2\,,\quad D_{10}=(l_1 n)\,,\nonumber\\
&D_{11}=(l_2 n)\,,\quad D_{12}=(l_1-q_2)^2  \,,\quad D_{13}=(q_1 n)\,,\quad D_{14}=(q_2 n)\,.
\label{eq:Ds}
\end{align}
Here $n=(1,\boldsymbol{0})$ is a time-like unit vector and we put $m=1$.
The functions $D_{1-9}$ and $D_{13-14}$ correspond to the denominators of the topology depicted in Fig. \ref{fig:top}.
The $\delta$-functions in Eq. \eqref{eq:j} secure the energy conservation. Note that $n_{10-12}\leqslant 0$ and the  prescription $-i 0$ for $D_{1-9}$ is implied.

\begin{figure}[h]
	\centering
	\includegraphics[width=1.\linewidth]{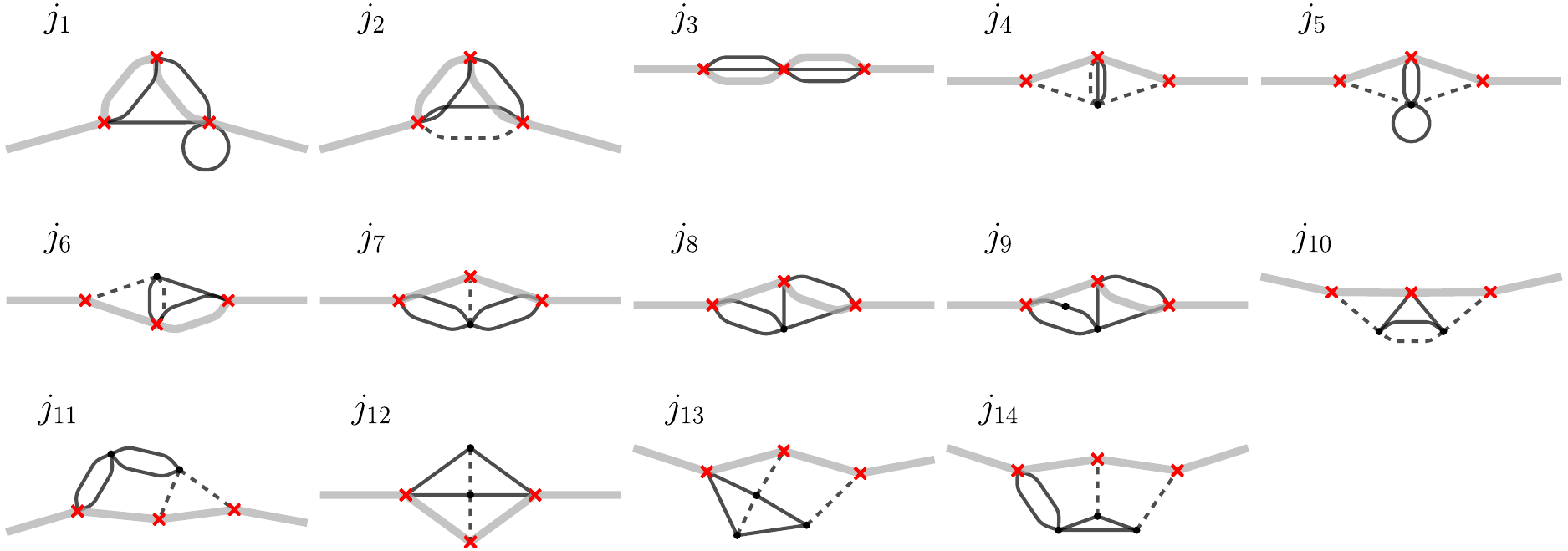}
	\caption{Master integrals.}
	\label{fig:MIs}
\end{figure}

Making the IBP reduction \cite{Chetyrkin1981,Tkachov1981}  with \texttt{LiteRed} \cite{Lee2014},  we reveal $14$ master integrals, see Fig.~\ref{fig:MIs}.

Note that the counter-term diagrams in the last line of Fig. \ref{fig:diagrams} can also be expressed in terms of the four-loop master integrals in Fig. \ref{fig:diagrams} although they have only three loops. To this end we multiply the corresponding integrals by $1=\frac{-1}{\Gamma[1-d/2]}\int \frac{d^d l_2}{i\pi^{d/2}D_2}$. Then the contribution of counter-terms is expressed via the master integrals with unit mass tadpole loop, namely, via $j_1$ and $j_5$.

\begin{figure}[h]
	\centering
	\includegraphics[width=1.\linewidth]{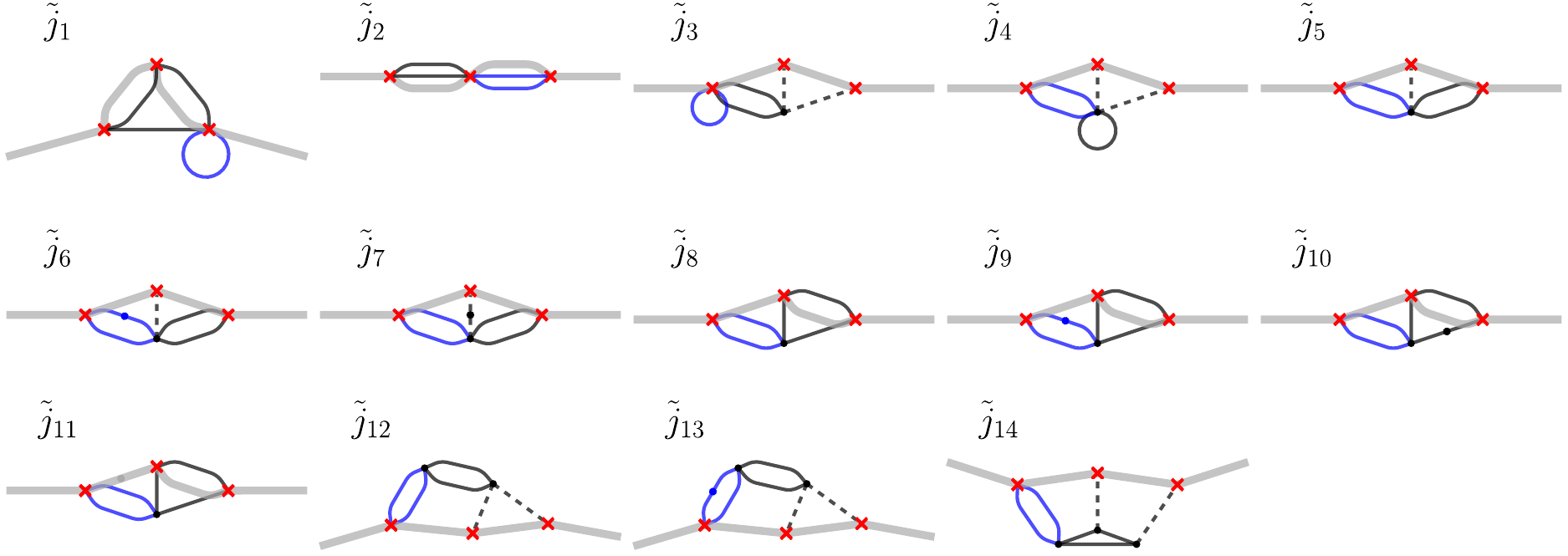}
	\caption{Master integrals of the two-scale family. Blue lines denote massive propagators with mass $m$.}
	\label{fig:mMIs}
\end{figure}
Since the integral family \eqref{eq:j} contains no dimensionless free parameter, the differential equations method can not help directly. Therefore, there is a temptation to calculate the master integrals with the DRA method \cite{Lee2010}. Unfortunately, there is a nontrivial $2\times 2$ diagonal block in the matrix of dimensional recurrence, corresponding to the integrals $j_8$ and  $j_9$ which belong to one and the same sector. Although it is, in principle, possible to apply the DRA method also for the cases with nontrivial diagonal $2\times 2$ blocks as discussed in Ref. \cite{Lee:2017ftw,Lee:2017yex}, its application in this case is much more laborious than that for the triangular matrix. Fortunately, for the present task we may apply a combined approach. First, we use the DRA method for all integrals but $j_8$,  $j_9$, and $j_{14}$ (the latter integral belongs to a super-sector of the sector of $j_{8,9}$). In order to obtain information about analytical properties necessary for fixing periodic functions in homogeneous parts of the solutions, we use the approach of Ref. \cite{Lee:2022art}. Namely, we choose the integrals which are obviously finite on a sufficiently wide vertical stripe in the complex plane of $d$. For example, in order to reveal the analytic properties of the most complicated integral $j_{13}$ we use finiteness of the integral $j_{13f}=\raisebox{-2mm}{\includegraphics[width=2cm]{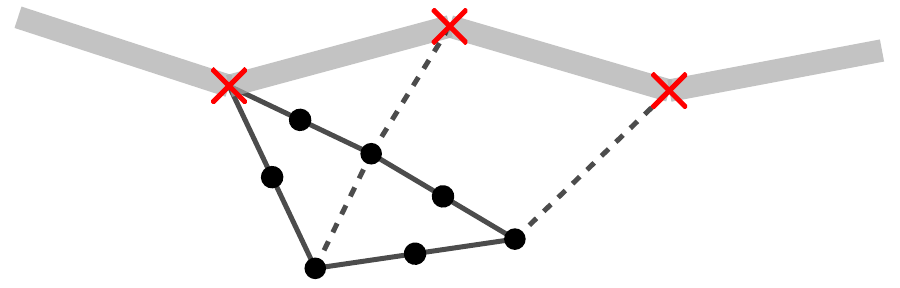}}$ on the stripe $\Re d\in (3,5]$. Reducing $j_{13f}$ to the master integrals, we obtain a number of nontrivial constraints for the leading expansion terms of $j_{13}$ on  the basic stripe $(3,5]$.
The results of the DRA approach have the form of $n$-fold triangular sums with factorized summand and $n\leqslant 3$, We use \texttt{SummerTime} package \cite{Lee:2015eva} to calculate the $\epsilon$-expansion of these sums with high precision and PSLQ algorithm \cite{ferguson1999analysis} to recognize the result in terms of multiple zeta values.

Note that all three remaining integrals, $j_8$,  $j_9$, and $j_{14}$,  contain two disjoint massive loops. In order to calculate those integrals, we consider a family of integrals with different masses in these two loops. This family is defined by  Eqs. \eqref{eq:j} and \eqref{eq:Ds} where one should replace $D_1\to \widetilde{D}_1=m^2-l_1^2$, $D_5\to \widetilde{D}_5 = m^2-(l_1+q_1)^2$ and assume that $n_{6,7}\leqslant 0$ (in addition to $n_{10-12}\leqslant 0$). This family corresponds to the denominators of $j_{14}$, where the unit mass in the left-most fermion loop is replaced by $m$. Performing the IBP reduction, we reveal 14 master integrals depicted in Fig. \ref{fig:mMIs}. We obtain the differential system
\begin{equation}\label{eq:DE}
	\partial_m\tilde{j} = M(\epsilon,m) \tilde{j}
\end{equation}
for the column $\tilde{j}=(\tilde{j}_1,\ldots,\tilde{j}_{14})^\intercal$ and reduce it to $\epsilon$-form using \texttt{Libra} \cite{Lee:2020zfb}. The general solution of Eq. \eqref{eq:DE} is expressed in terms of harmonic polylogarithms \cite{Remiddi:1999ew}. The boundary conditions are put at $m\to 0$. We use expansion by regions \cite{Beneke:1997zp,Pak:2010pt} to fix the required coefficients in the asymptotic expansion of integrals.
The original integrals $j_{8,9,14}$ are recovered as
\begin{equation}
	j_8=\left.\tilde{j}_8\right|_{m=1},\quad
	j_9=\left.\tilde{j}_9\right|_{m=1},\quad
	j_{14}=\left.\tilde{j}_{14}\right|_{m=1}.
\end{equation}
Note that other integrals in Fig. \ref{fig:MIs} which contain two disjoint fermion loops are also expressed in terms of the master integrals in Fig. \ref{fig:mMIs}:
\begin{equation}
	j_1=\left.\tilde{j}_1\right|_{m=1},\quad
	j_3=\left.\tilde{j}_2\right|_{m=1},\quad
	j_4=\left.\tilde{j}_3\right|_{m=1}=\left.\tilde{j}_4\right|_{m=1},\quad
	j_7=\left.\tilde{j}_5\right|_{m=1},\quad
	j_{11}=\left.\tilde{j}_{12}\right|_{m=1}.
\end{equation}
This provides a number of nontrivial cross checks of the obtained results for the master integrals.

Finally, we expand the diagrams in Fig. \ref{fig:diagrams} in $q$ and perform the Dirac algebra using \texttt{FeynCalc} \cite{Shtabovenko:2020gxv}. After the IBP reduction and substitution of the results for the master integrals, we obtain our final result for the coefficient $B$ in Eq. \eqref{eq:deltaE}. We present the contributions of diagrams in Figs. \ref{fig:ph} and \ref{fig:po} separately:
\begin{align}
B_{\ref{fig:ph}}&=\frac{1456}{45} \mathrm{Li}_4\big(\tfrac{1}{2}\big)
	-\frac{4511 \pi ^4}{16200}
	+\frac{182 \ln^4{2}}{135}
	+\frac{274}{135} \pi ^2 \ln^2{2}
	-\frac{2387 \pi^2 \ln{2}}{1080}
	+\frac{199 \zeta_3}{45}\nonumber
	\\
	&
	+\frac{13057}{3240}
	+\frac{3703 \pi^2}{5760}=0.125181281880322\ldots\,,
	\\
B_{\ref{fig:po}}&=\frac{71 \zeta_3}{56}-\frac{479}{756}+\frac{38401 \pi ^2}{217728}-\frac{283}{756} \pi ^2 \ln{2} = 0.070271202837585\ldots\,,
\\
B&=B_{\ref{fig:ph}}+B_{\ref{fig:po}}=0.195452484717907\ldots\,.
\label{eq:result}
\end{align}

\section{Conclusion}

In the present paper we obtain the  contributions of order $\alpha^2(Z\alpha)^6m$ to the Lamb shift from radiative corrections to the Wichmann-Kroll potential depicted in Fig.~\ref{fig:diagrams}. Numerically our result \eqref{eq:result} appears to be rather small and compatible with the heuristic estimate $B=0.13\pm 0.13$ of Ref. \cite{Karshenboim2019}. For the calculation of master integrals we use a combination of the DRA method and the approach based on the differential equations. This calculation provides yet another example of the effectiveness of multiloop methods for obtaining analytic results in atomic physics.

\section*{Acknowledgments}
The work has been supported by Russian Science Foundation under grant 20-12-00205.

\bibliographystyle{JHEP}
\bibliography{WK}
\end{document}